# Hybrid Graphene-Plasmonic Gratings to Achieve Enhanced Nonlinear Effects at Terahertz Frequencies

Tianjing Guo, Boyuan Jin, and Christos Argyropoulos[*]
Department of Electrical and Computer Engineering, University of Nebraska-Lincoln, Lincoln, Nebraska 68588, USA

[*]christos.argyropoulos@unl.edu

High input intensities are usually required to efficiently excite optical nonlinear effects in ultrathin structures due to their extremely weak nature. This problem is particularly critical at low terahertz (THz) frequencies because high input power THz sources are not available. The demonstration of enhanced nonlinear effects at THz frequencies is particularly important since these nonlinear mechanisms promise to play a significant role in the development and design of new reconfigurable planar THz nonlinear devices. In this work, we present a novel class of ultrathin nonlinear hybrid planar THz devices based on graphene-covered plasmonic gratings exhibiting very large nonlinear response. The robust localization and enhancement of the electric field along the graphene monolayer, combined with the large nonlinear conductivity of graphene, can lead to boosted third harmonic generation (THG) and four-wave mixing (FWM) nonlinear processes at THz frequencies. These interesting nonlinear effects exhibit very high nonlinear conversion efficiencies and are triggered by realistic input intensities with relative low values. In addition, the THG and FWM processes can be significantly tuned by the dimensions of the proposed hybrid structures, the doping level of graphene, or the input intensity values, whereas the nonlinear radiated power remains relatively insensitive to the incident angle of the excitation source. The presented nonlinear hybrid graphene-covered plasmonic gratings have a relative simple geometry and, as a result, can be used to realize efficient third-order nonlinear THz effects with a limited fabrication complexity. Several new nonlinear THz devices are envisioned based on the proposed hybrid nonlinear structures, such as frequency generators, all-optical signal processors, and wave mixers. These devices are expected to be useful for nonlinear THz spectroscopy, noninvasive THz subwavelength imaging, and THz communication applications.

## I.   Introduction

Graphene is a two-dimensional (2D) material with unique electric and optical properties [1,2]. Surface plasmons are formed at its surface when excited by THz radiation [3], a property of paramount interest for a material to be used in the envisioned integrated THz plasmonic systems [4]. In addition, the conductivity of graphene can be dynamically controlled and tuned by electrostatic doping, usually by using a gating voltage configuration [5]. The tunable and reconfigurable functionalities of graphene have recently been widely investigated with the goal to design different adaptive graphene-based THz devices, such as polarizers [6], cloaks [7], phase shifters [8], optical modulators [9], and absorbers [10]. However, light absorption along a graphene monolayer is usually very weak due to its single-atom thickness, which consists a deleterious property towards the practical applications of graphene-based devices [11]. Fortunately, the absorption can be enhanced by patterning doped graphene monolayers into periodic nanodisks [12], combining graphene with insulating layers [13], or integrating graphene with microcavities

[14]. Nevertheless, complicated fabrication procedures are required to fabricate most of the aforementioned graphene-based configurations making these designs prone to fabrication imperfections and other limitations. Recently, the magnetic resonance of a metallic grating was used to enhance the absorption of graphene monolayers at THz frequencies [15]. This structure is easier to be fabricated and can potentially be more practical towards the design of compact THz devices. By placing a graphene monolayer over a metallic grating, strong and localized electric fields are obtained along the graphene when the plasmon modes of both graphene and grating coincide. Note that this grating design is different compared to the recently proposed hybrid plasmonic waveguide modulator loaded with graphene [16].

Interestingly, graphene has been found to possess strong nonlinear electromagnetic properties [17,18]. The second-order nonlinear response of a graphene monolayer usually vanishes within the dipole approximation [19], since graphene has centrosymmetric properties. However, graphene has been experimentally demonstrated to possess a remarkably strong third-order nonlinear susceptibility $\chi^{(3)}$ at THz frequencies [20]. The strong third-order nonlinear response originates from the intraband electron transitions [21], as well as the resonant nature of the light-graphene interactions. Both of these effects are dominant under THz radiation illumination. Specifically, the Kerr nonlinear susceptibility $\chi^{(3)}$ of graphene was found to reach high values $\left(1.4 \times 10^{-15} \, m^2/V^2\right)$ in recent experiments [22].

Two of the most common third-order nonlinear processes are third-harmonic generation (THG) and four-wave mixing (FWM). In the case of THG, an incident wave $(\omega)$ interacts with the system to produce a wave with three times the incident wave frequency $(3\omega)$ [23]. The significant advantage of THG compared to other nonlinear processes is that it can be achieved by using a single-wavelength source. This nonlinear process can be used to realize higher transverse resolution for nonlinear imaging and microscopy techniques [24] and improved sensing [25]. THG has also been reported to be generated by graphene and few-layer graphite films but with relative low efficiency despite the large nonlinear graphene susceptibility [26,27]. FWM is another interesting third-order nonlinear process that has found a plethora of applications in nonlinear imaging, wavelength conversion, optical switching, and phase-sensitive amplification [28–32]. Different to THG, during the process of FWM, two pump $\omega_1$ and one probe $\omega_2$ photons are absorbed and mixed at the nonlinear medium and a photon at $\omega_{FWM} = 2\omega_1 - \omega_2$ is generated and re-emitted. The FWM efficiency strongly depends on the field enhancement at the input pump and probe frequencies and proper phase matching conditions [33]. The field enhancement along graphene and other 2D materials is usually very weak due to the poor coupling of the incident electromagnetic radiation to these ultrathin media. However, the phase matching condition can be relaxed in the case of 2D materials, since they are extremely thin and phase cannot be accumulated along their thickness in contrast to conventional bulk nonlinear materials. The increase in the nonlinear efficiency of 2D materials still remains elusive and is subject of intense on-going research [34, 35].

Usually, very high input intensities are required to excite third-order nonlinear effects from ultrathin nonlinear materials due to their extremely weak nature leading to very poor nonlinear efficiency [36]. This detrimental effect is particular acute at low THz frequencies, since high input power THz sources do not exist [37]. The use of different plasmonic configurations has been demonstrated to efficiently boost optical nonlinear effects mainly due to the enhanced local electric fields and relaxed phase matching conditions at the nanoscale [38–52]. Plasmonic effects can indeed lead to enhanced effective nonlinear susceptibilities based on different configurations which are composed of materials with weak intrinsic nonlinear properties. Yet, the plasmonic boosting of nonlinear effects has been mainly achieved in the infrared and visible spectrum and the enhancement of nonlinearities at THz frequencies still remains limited.

In the current work, we study an alternative approach to boost nonlinear effects specifically at THz frequencies based on ultrathin hybrid plasmonic structures. In particular, we investigate the potential of graphene-covered metallic gratings to dramatically amplify the inherently weak nonlinear response of conventional metallic gratings and isolated 2D materials [34]. More specifically, it will be demonstrated that the addition of graphene in the proposed hybrid plasmonic structures is of paramount importance to the boosting of different nonlinear procedures. This is mainly due to three reasons: a) the large field enhancement and confinement arising from the strong interference between the surface plasmon excited due to the graphene monolayer and the localized plasmon confined in the metallic grating [53]. b) The high nonlinear response of graphene [54], which is further boosted in the presence of the strong localized fields despite its one-atom thickness. c) The perfect phase matching condition that is achieved along the subwavelength thickness of graphene. Due to these reasons, the efficiency of both THG and FWM processes will be greatly enhanced with the proposed graphene-covered grating structures. Furthermore, the third-order nonlinear graphene conductivity is a function of the Fermi energy or doping level of graphene [55]. This effect will be utilized to dynamically control the presented THz third-order nonlinear processes by tuning the doping level via a gate voltage configuration [3], [6]. The voltage values required for the aforementioned tuning are relative low, as will be demonstrated later. This tunable mechanism allows the efficient generation of enhanced reconfigurable THz nonlinearities and provide the possibility to develop adaptive graphene-based nonlinear THz devices [56–60].

In order to investigate the enhancement in the nonlinear response of the proposed graphene-covered plasmonic grating, we employ full-wave numerical simulations based on the finite element method (FEM) by using COMSOL Multiphysics. The FEM equations are substantially modified to include the appropriate nonlinear polarizabilities and currents in Maxwell's equations in order to be able to accurately simulate the presented nonlinear effects. The presented full-wave simulations are ideal to calculate the nonlinear radiated power by integrating the power outflow over a surface that surrounds the device under study. This consists ideal simulation scenario to accurately compare the obtained theoretical results with the potential experimental results that will be attained by nonlinear experiments. The proposed hybrid structure can be easily experimentally verified with conventional fabrication techniques. For example, the graphene and metallic grating can be separately fabricated, using chemical vapor deposition on a copper foil and e-beam lithography [61, 62], respectively, and then combined to create the proposed hybrid structure. Note that the presented plasmonic grating and graphene can only support plasmons when they are excited by transverse magnetic (TM) polarized waves [63] in the currently used THz frequency range. Hence, TM polarized incident waves are considered as the excitation source of the proposed system throughout this work.

The paper is organized as follows: first, we theoretically analyze the linear response of the proposed hybrid graphene-metal system by using its equivalent circuit model. The validity of the theoretical method is verified by comparing several analytical results with full-wave simulations in Section II. The effects of the geometrical dimensions, graphene Fermi level, and different incident angles are also investigated in the same section. Next, we present in Sections III and IV the enhancement of the THG and FWM nonlinear processes, respectively, due to the proposed hybrid nonlinear graphene-covered plasmonic grating. By comparing the nonlinear performance of the proposed hybrid gratings to other conventional non-hybrid structures, such as a bare graphene monolayer or a plain metallic grating, it is demonstrated that the efficiencies of both THG and FWM are increased by multiple orders of magnitude. Moreover, it is shown that the presented enhanced nonlinear responses can become tunable by varying the geometry of the proposed hybrid grating or the graphene doping level. Finally, we also demonstrate the relative insensitivity of the proposed system to the angle of incidence when oblique incident illumination is considered.

## II.   Perfect Absorption of THz Radiation

The geometry of the proposed hybrid graphene-covered metallic grating is illustrated in Fig. 1(a). The grating is periodic in the x-direction with period $p$ and is assumed to be extended to infinity in the y-direction. It is made of gold (Au) with THz optical constants calculated by using the Drude model: $\varepsilon_{L,Au} = \varepsilon_\infty - f_p^2/f(f-i\gamma)$, where $f_p = 2069\,THz$, $\gamma = 17.65\,THz$ and $\varepsilon_\infty = 1.53$ are derived from fitting the experimental values [64]. The height and trench width of the grating are equal to $d$ and $b$, respectively. The ground plane is thick enough to be considered opaque to the impinging THz radiation leading to zero transmission. During our theoretical analysis, we initially assume the metallic grating to be made of perfect electric conductor (PEC), which is a good approximation since the electromagnetic fields minimally penetrate metals at low THz frequencies [65]. The grating is covered by a graphene monolayer sheet with doping or Fermi level $E_F$. Intraband transitions dominate the graphene response in the low THz frequency range and the linear conductivity of graphene can be expressed by using the Drude formalism: $\sigma_g = \dfrac{e^2 E_F}{\pi \hbar^2} \dfrac{j}{j\tau^{-1} - \omega}$ [66], where $e$ is the electron charge, $\hbar$ is the reduced Planck's constant, $\omega = 2\pi f$ is the angular frequency, and $\tau$ is the relaxation time, which is assumed to be equal to $\tau = 10^{-13}\,s$ throughout this work.

The proposed structure is illuminated by a TM polarized wave (the magnetic field in the y-direction) with an incident angle $\theta$ with respect to the z-direction. The inset at the left side of Fig. 1(b) shows the equivalent circuit [67] used to analytically model one unit cell of the graphene-covered grating with $Y_1 = \omega \varepsilon_0 n / k_0$ and $Y_2 = \omega \varepsilon_0 n / k_0 (p/b)$ being the corresponding characteristic admittances of the surrounding air and grating trench regions, respectively. In these formulas, $n$ is the refractive index of air around the grating and inside the grating trench and $k_0$ is the free space wavenumber. The graphene sheet is modeled as an additional shunt admittance $Y_S$ in the equivalent circuit model that can be calculated by the simple formula: $Y_S = \sigma_g (p/b)$. The reflection coefficient is computed by applying the transmission line theory [68] to the proposed THz equivalent circuit model and is equal to: $\Gamma = \dfrac{Y_1 - [Y_S - jY_2 \cot(\beta d)]}{Y_1 + [Y_S - jY_2 \cot(\beta d)]}$ [69], where $\beta$ is the propagation constant in the grating trench region. The absorptance can be computed by the relationship $A = 1 - |\Gamma|^2$ [12, 15, 70-72], since in the current grating configuration the transmission is equal to zero. Alternatively, the total reflected power density from the hybrid grating can be used to compute the absorptance by using the formula:

$absorptance = 1-(reflected\ power/incident\ power)$, which will lead to similar results. The absorptance versus the frequency, computed by using the proposed transmission line model, is plotted in Fig. 1(b) (black line) at normal incident illumination $(\theta = 0°)$ for a grating with dimensions *p=10um, b=0.6um, d=10um* loaded with graphene on top with Fermi level 0.3eV. Note that the trench width *b* is much smaller than the period *p* in all our designs and the graphene can be placed over the grating without being bended at the corrugations.

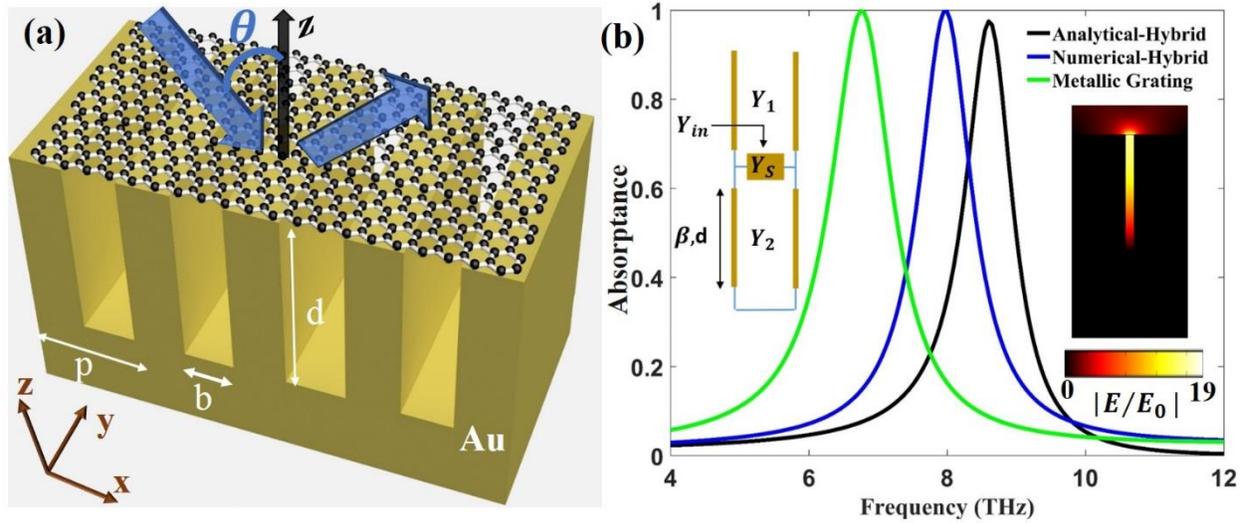

Figure 1 - (a) Schematic of the hybrid graphene-covered gold grating. (b) Analytically (black line) and numerically (blue line) computed absorptance spectra of the proposed hybrid grating. The absorptance of the metallic grating without graphene on top is also plotted with the green line. The results are obtained for grating parameters *p=10um, b=0.6um, d=10um*. The graphene's Fermi level $E_F$ is equal to $0.3eV$. The left inset shows the equivalent circuit model used to theoretically analyze the proposed structure. The right inset represents the computed electric field enhancement distribution at the resonant frequency of the hybrid grating.

In order to verify the accuracy of the presented equivalent circuit model, we also compute the response of the proposed structure by using numerical simulations based on COMSOL Multiphysics. The structure is again assumed to be infinite along the y-direction in the numerical modeling case, as shown in Fig 1(a), and is modeled as a 2D system to accelerate the calculations. Periodic boundary conditions are employed in the x-direction and port boundaries are placed in the z-direction to create the incident plane wave. Graphene is modeled as a surface current, due to its planar (2D) nature, described as $J = \sigma_g E$, where $E$ is the electric field along its surface and $\sigma_g$ is the linear graphene conductivity given before by the Drude model.

The computed absorptance based on both theoretical and numerical methods are shown in Fig. 1(b) with black and blue lines, respectively, and are found to be in good agreement. The small frequency shift between the theoretical and simulation results can be attributed to the approximation of gold with PEC in the theoretical model, as well as to the used finite size mesh during the full-wave modeling. However, both results are very similar and one pronounced perfect absorptance peak is demonstrated in Fig 1(b) at the resonance of the hybrid grating. At this resonant point, a magnetic plasmon mode is formed due to the generation of highly localized magnetic fields inside the grating's trench accompanied by high electric fields that are expected to boost nonlinearities [15]. The electric field enhancement distribution along the structure at the resonant frequency is demonstrated by the right inset in Fig. 1(b) which is computed by calculating the ratio $|E/E_0|$, where $E_0$ represents the amplitude of the incident electric field. Interestingly, the electric field can be enhanced by approximately nineteen times inside the grating's trench and, more importantly, along its surface, where graphene will be deposited. The obtained perfect absorption indicates a strong coupling and interference between the THz plasmons of graphene and metallic grating. In addition, we calculate the absorptance of a plain metallic grating (i.e., without graphene on top of it) by using the same numerical method. The result is shown by the green line in Fig. 1(b). It is interesting that a substantial frequency blueshift is obtained when graphene is introduced over the grating. The addition of graphene will also lead to dynamic tunability in the absorption resonant frequency, as will be demonstrated later. When the polarization of the incident wave is switched and the magnetic field is oriented along the length of the grating (transverse electric (TE) polarization), no resonance is observed since both the graphene and metallic grating cannot support TE plasmons and, as a result, cannot couple to the incident TE radiation. We also verified that the absoprtance of a flat gold substrate with and without graphene on top is very low because there is no plasmon formation along the flat interface [73].

Next, we investigate the effect of the grating's geometry to the calculated perfect absorption of the proposed hybrid graphene grating. Throughout this work and only if otherwise specified after this point, we always assume the following realistic graphene parameters: $E_F = 0.1 eV$, $\tau = 10^{-13} s$ and the following practical to realize grating microscale dimensions: *p=8um, d=8um,* and *b=0.6um*. Figure 2(a) displays a contour plot of the computed absorptance for TM polarized plane waves impinging at normal incidence as a function of the frequency and period, where the grating's dimensions *d=8um* and *b=0.6um* have fixed values. Clearly, the absorptance remains strong as we vary the period p from 5um to 25um, indicating a strong coupling between grating and graphene plasmons independent of the periodicity. We have also verified that higher-order diffraction modes or surface waves are not excited by the proposed grating if the incident wave has frequencies in the currently used range of 6-12 THz. The resonant frequency of the perfect absorptance is slightly affected by the period and the bandwidth of the resonance peak is decreased as the period is increased. Interestingly, the resonant frequency of the perfect absorption is found to be more sensitive to the height of the grating. It is decreased as the grating height is increased, as illustrated in Fig. 2(b), where the computed absorptance is plotted as a function of the frequency and grating height. Thus, it is possible to tune the perfect absorption to different frequencies by modulating the grating geometry.

The perfect absorptance can also be tuned without changing the grating geometry but by electrostatically gating the graphene leading to a change in its doping level. The effect of different graphene Fermi level values on the perfect absorptance of the proposed graphene-covered gratings is demonstrated in Fig. 2(c). There is a substantial shift in the resonant frequency of the perfect absorptance as the Fermi level is increased. The modification in the Fermi level leads to different graphene properties and, as a result, to a frequency shift in the resonant response of the graphene plasmons. However, it is interesting that the absorptance remains perfect between 0.1 eV to 0.45 eV and only blueshifts as the doping level is increased. This doping variation can be achieved by electrostatically gating the graphene monolayer with a pair of transparent electrodes, as explained in the next paragraph [74]. Finally, we investigate the performance of the proposed graphene-covered grating under different incident angles of the excitation wave. The calculated absorptance is shown in Fig. 2(d) as a function of the incident angle and frequency. Evidently, the absorptance remains almost perfect located at the same resonant frequency point over a wide range of incident angles, in particular between ±80°.

The currently proposed structure can be realized with existing well-established fabrication methods, since just a graphene monolayer needs to be deposited on a microscale metallic grating. The gating voltage can be applied only on the suspended part of the graphene sheet because the remaining graphene part will be shorted while touching the metallic grating. Note that the portion of graphene along the grating ridges has no effect on the absorption and nonlinear response of the proposed hybrid structure. Hence, the nonuniform doping profile of graphene will not affect the response of the presented configuration. This is due to the fact that only the suspended part of graphene over the trenches strongly interacts with the incident absorbed power, as it is clearly shown by the field and power profiles plotted in the Supplemental Material [73]. Surface waves are not excited along the grating ridges at the perfect absorption frequency and only localized power is formed on the upper part of the trenches at this frequency point, as it is depicted in the Supplemental Material [73]. As a result, the fields along the trenches of the grating are the strongest [73], where the graphene monolayer is located. Thus, the nonlinear signal will be mainly generated by the strong fields interacting with graphene in these nanoregions. The metallic trenches can be loaded with a dielectric material to efficiently gate the suspended graphene sheet leading to a voltage-controlled perfect absorber. The relationship between Fermi level $E_F$ and gate voltage $V_g$ for this gating configuration is: $E_F = hv_F\sqrt{\pi CV_g/e}$ [75], where $h$ is the Planck constant and $v_F = 1\times 10^6\ m/s$ is the graphene Fermi velocity. The formed electrostatic capacitance per unit area $C$ is equal to $C = \varepsilon_0\varepsilon_d/d$, where $\varepsilon_d$ and $d$ are the dielectric permittivity and thickness (height) of the grating trench, respectively. Here, we assume $\varepsilon_d = 4.4$ and the highest gate voltage value to achieve the maximum used Fermi level (0.45 eV) is computed to be 123 V, which is realistic and relative low value paving the way towards a potential experimental verification of the proposed tunable THz absorber [76].

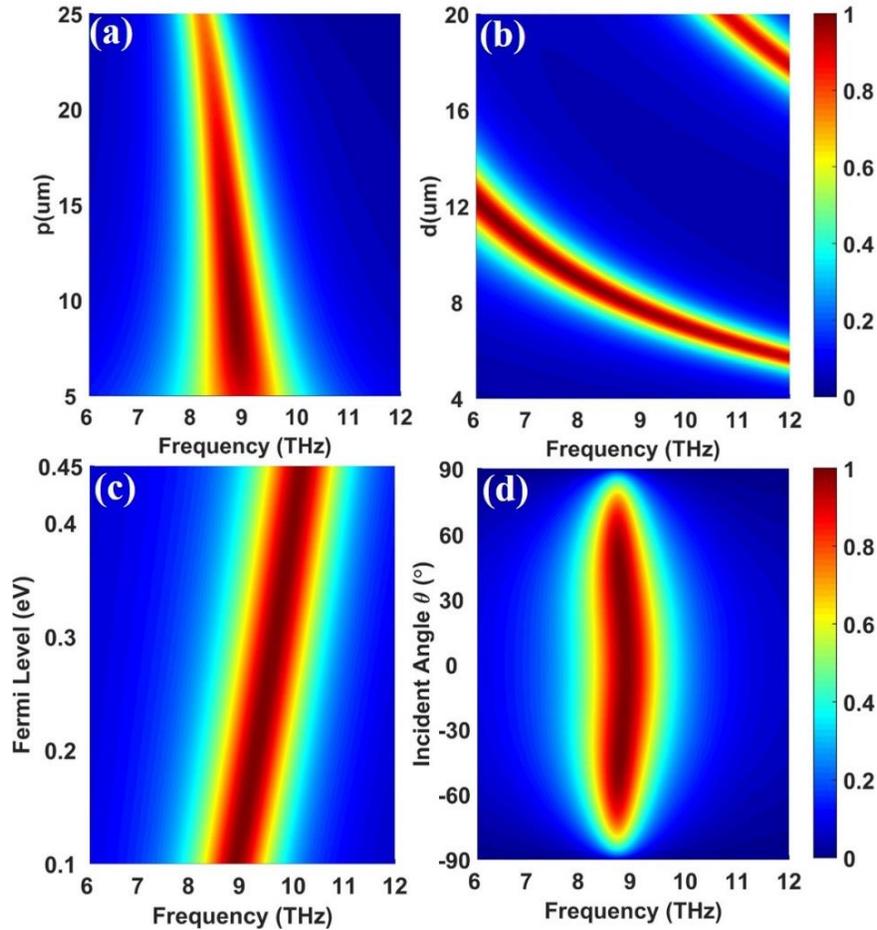

Figure 2 - Computed absorptance contour plots of the graphene-covered grating as a function of the frequency and (a) period $p$ and (b) height $d$ of the grating, (c) Fermi level of graphene, and (d) incident angle of the excitation wave.

## III. Third Harmonic Generation

During the analysis presented in the previous Section II, we proved that perfect and tunable absorption can be achieved by using a hybrid metallic grating covered by graphene. It was demonstrated that the electric field is greatly enhanced at the absorption resonance due to the strong coupling between the graphene and grating plasmonic responses. The increased light-matter interactions achieved by the proposed hybrid structure and obtained at the perfect absorptance frequency point have the potential to dramatically enhance the nonlinear response of graphene at low THz frequencies. Towards this goal, in this section, we investigate the THG efficiency of the proposed graphene-covered grating when all the nonlinear properties of the used materials are included in our nonlinear simulations. The fundamental frequency (FF) that excites the nonlinear system is always set to coincide with the perfect absorptance resonant frequency of the proposed hybrid structure. The strong electric fields at the resonance will subsequently boost the excited nonlinear effects.

Gold can be assumed to exhibit PEC-like response at low THz frequencies, since it has very high conductivity and the fields minimally penetrate its bulk volume. However, in order to ensure the accuracy of our nonlinear simulations, we include its nonlinear susceptibility $\chi_{Au}^{(3)} = 2.45\times 10^{-19}\ m^2/V^2$ at the infrared frequency region [77], in addition to its linear Drude model, as it was described in the previous section. Hence, the Kerr nonlinear permittivity of gold is given by: $\varepsilon_{NL,Au} = \varepsilon_{L,Au} + \chi_{Au}^{(3)}E_{FF}^2$, where $E_{FF}$ is the enhanced electric field induced at the FF and shown in the right inset of Fig. 1(b). We will demonstrate later that the nonlinear response of the proposed system is dominated by the nonlinear properties of graphene and not the gold grating nonlinear permittivity.

The third-order nonlinear surface conductivity of graphene at THz frequencies is calculated by the formula [55]:

$$\sigma^{(3)} = \frac{i\sigma_0(\hbar v_F e)^2}{48\pi(\hbar\omega)^4}T(\frac{\hbar\omega}{2E_F}), \qquad (1)$$

where $\sigma_0 = e^2/4\hbar$, $v_F = 1\times 10^6\ m/s$, $T(x) = 17G(x) - 64G(2x) + 45G(3x)$, with $G(x) = \ln|(1+x)/(1-x)| + i\pi\theta(|x|-1)$, and $\theta(z)$ is the Heaviside step function. Graphene is modeled in COMSOL as a nonlinear surface current that can be expressed as $J = \sigma_g E_{TH} + \sigma^{(3)}E_{FF}^3$

[78], where $E_{FF}$ and $E_{TH}$ are the electric field induced at the FF and the third harmonic (TH), respectively. An additional electromagnetic wave solver needs to be included in COMSOL and coupled to the FF solver in order to accurately compute the THG radiation, which will solve the nonlinear Maxwell's equations at the TH frequency $\omega_{TH} = 3\omega$. The surface current formalism used to model the nonlinear graphene leads to more accurate simulation results combined with less stringent mesh quality requirements. This type of simulations are faster and more accurate compared to the widely used conventional numerical three-dimensions (3D) graphene modeling, where graphene is considered to be a bulk material [59]. The undepleted pump approximation is used in all our nonlinear simulations, since the nonlinear signals are expected to be weaker compared to their linear counterparts.

The schematic of the THG process is illustrated in Fig. 3(a). In this case, a wave with frequency $3\omega$ will be generated when an incident wave with FF $\omega$ excites the proposed nonlinear graphene-covered grating. The insets in Fig. 3(b) demonstrate the electric field enhancement distributions for this structure at FF and TH frequencies, respectively. Clearly, enhanced localized electric fields are obtained both at the FF resonance, as well as the TH frequency. In order to take advantage of the strong field enhancement and boost THG, the frequency of the FF excitation wave is located close to the perfect absorption resonance which is computed to be around $f = 8.8 THz$ in this case. The proposed nonlinear structure is illuminated at normal incidence with a relative low input intensity equal to $10 kW/cm^2$. This value is substantially lower compared to intensity values used in previous works based on just the nonlinear properties of graphene without the addition of plasmonic structures [27]. By calculating the integral $\int_C \vec{S} \cdot \vec{n}$ over the entire surface of the structure and at the far field, where $\vec{S}$ is the Poynting vector crossing the boundary surface C and $\vec{n}$ is the boundary norm vector, the radiated output power of the TH wave with radiation frequency $3f = 26.4 THz$ is computed. Note that we assume a fictitious large distance of one meter for the y-direction length of the proposed structure in all our 2D simulations, thus the output power is always computed in Watt units. The result of the radiated output power of the TH wave is shown in Fig. 3(b). Interestingly, there is a peak at the TH radiation around the absorption resonance, which coincides with the FF, and its maximum value can reach 1.6 W. In addition, the TH radiation power shown in Fig. 3(b) follows the same trend with the absorptance shown before in Fig. 1(b). Finally, we would like to stress that reflected waves do not exist at the fundamental frequency $f = 8.8 THz$ due to the high absorptance of the proposed hybrid grating but there are strong THG reflected waves due to low absorptance at the third-harmonic frequency $3f = 26.4 THz$.

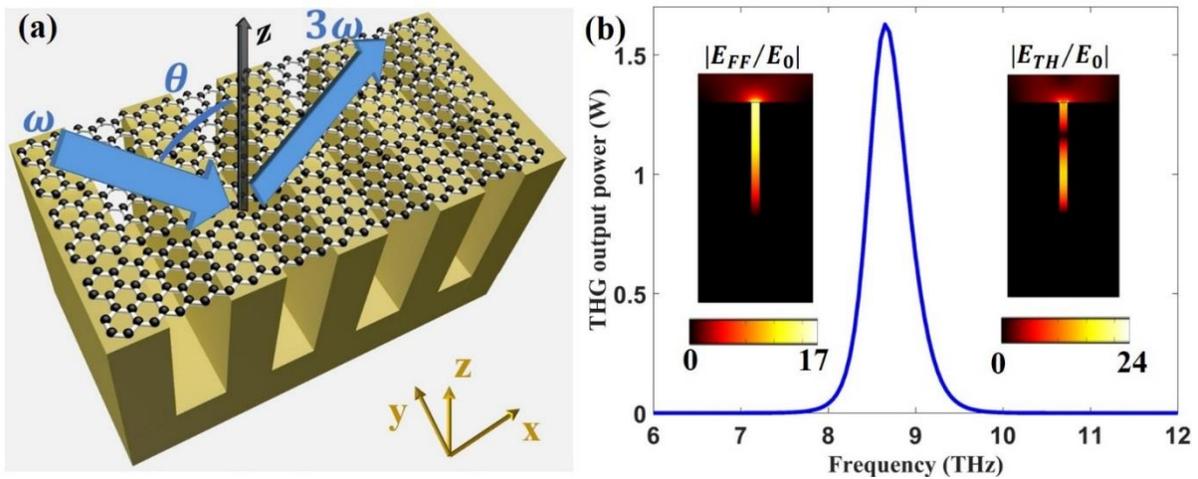

Figure 3 - (a) Schematic of the THG process due to the nonlinear graphene-covered grating. (b) The computed output power of the THG wave as a function of the incident wave fundamental frequency. Inset: The computed electric field enhancement distributions at the FF equal to 8.8 THz (left) and the TH at 26.4 THz (right).

Next, we compute the THG conversion efficiency (CE) which is a suitable metric to describe the THG power strength in a more quantitatively way. It is defined as the ratio of the radiated THG power outflow $P_{out\_TH}$ over the input FF power $P_{in\_FF}$: $CE = P_{out\_TH}/P_{in\_FF}$ [79]. It is clear that the incident FF wave intensity plays a key role in the CE. The CE is plotted in linear scale as a function of the FF wave intensity in Fig. 4. The FF is fixed at 8.8 THz, where the maximum linear absorption is achieved for the currently used geometrical parameters *p=8um, b=0.6um* and *d=8um*. The CE is dramatically enhanced by increasing the incident intensity of the FF wave. Notably, the CE can reach to high values (16 %) with very low input intensities ($100 kW/cm^2$), which is a substantial improvement compared to the previously proposed strong THG obtained by patterned nonlinear graphene metasurfaces at similar THz frequencies [80]. It is even more interesting that this high efficiency can be achieved by the currently proposed more realistic and easy to fabricate configuration. In the proposed structure, a graphene monolayer is used to obtain strong nonlinear response instead of patterned graphene microribbons that can suffer from detrimental edge loss effects at their discontinuities and other fabrication imperfections. The used input intensities have realistic values and even higher THz radiation intensities on the order of several $MW/cm^2$ have been reported in previous works with specialized configurations [81, 82]. We would like to stress that the presented THG efficiencies are on the order of a few percent and these values are higher compared to most nonlinear plasmonic devices presented so far [38]. Note that the computed CE relationship as a function of the input intensity has a quadratic shape (Fig. 4), an expected trend for THG nonlinear conversion efficiency. The input intensity of the FF wave will always be fixed to the low value of $20 kW/cm^2$ at all the following calculations unless otherwise specified.

According to Eq. (1), it is expected that a stronger nonlinear response will be obtained by using lower Fermi level values, i.e., less doped graphene. The THG CE will also be affected by the FF $(\omega)$ since the value of the nonlinear surface conductivity of graphene given by Eq. (1) is inversely proportional to $\omega$. As a consequence, stronger nonlinear response and higher CE are expected to be achieved at lower THz frequencies. However, the enhanced electric fields at the FF resonance will also affect the THG process. In order to verify how the THG CE will be affected by all these different parameters, the CE of the proposed nonlinear structure is computed by sweeping the graphene Fermi level and the fundamental frequency. The contour plot of this result is shown in Fig. 5. Clearly, the THG CE is decreased as the Fermi level is increased or in the case of off-resonance operation. The maximum CE value is obtained for approximately $f_{FF} = 8.8 THz$ and slightly doped graphene with Fermi level $E_F = 0.1 eV$. This trend is consistent with the absorptance analysis performed in the previous section. It is interesting that low doped graphene, which is easier to be produced, can lead to enhanced nonlinear effects based on the proposed configuration.

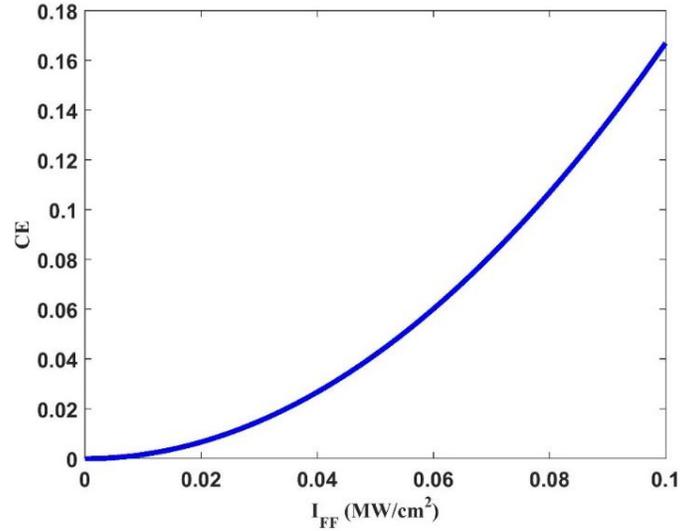

Figure 4 - Computed THG CE in linear scale as a function of the incident FF wave intensity. Relative low FF input THz wave intensities lead to very high THG CE.

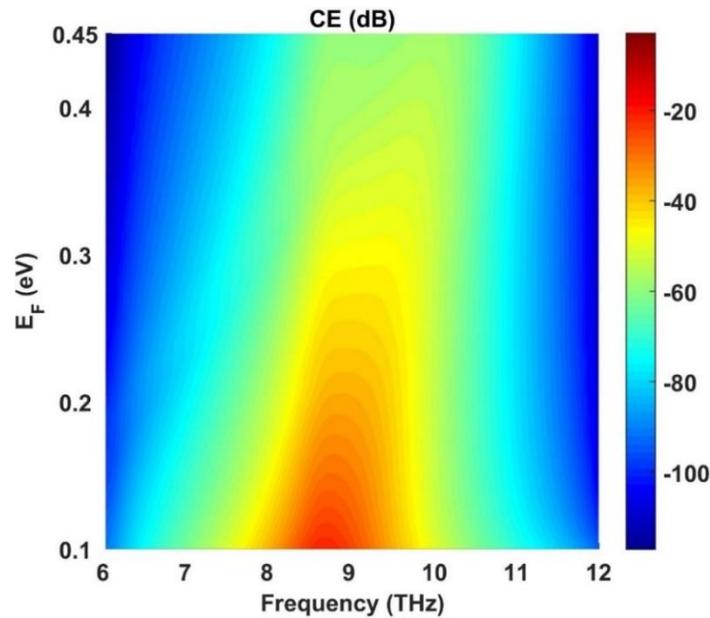

Figure 5 - The computed THG CE in logarithmic scale at normal incidence as a function of the fundamental frequency and the graphene's Fermi level of the proposed nonlinear hybrid graphene-covered grating.

We have discussed in Section II the effect of the proposed hybrid structure's geometry on the linear absorptance spectrum. In this section, we also investigate the effect of the geometry to the THG nonlinear process. Towards this goal, the THG CE is computed by sweeping the FF and the period or height of the grating, as shown in Figs. 6(a) and (b), respectively. The CE (plotted in dB) is tunable and follows the same trend with the linear absorptance enhancement illustrated before in Figs. 2(a) and (b). When the FF is tuned around the resonant frequency, a noticeable enhancement in the THG CE is observed for every period or height of the plasmonic grating with results shown in Fig. 6. This is directly related to the enhancement of the electric field at the absorptance resonance peak that boosts the nonlinear response of the structure translated to enhanced CE. In all the above simulations, we set the intensity of the illuminating wave to very low values $\left(20\,kW/cm^2\right)$, where the THG CE is relative high and equal to -22dB at the resonance. Hence, it is possible to also tune the THG nonlinear waves by changing the plasmonic grating's geometrical parameters and without altering the graphene properties. Note that there is no special requirement in the fabrication of the graphene monolayer used in the proposed hybrid structure since different values of graphene relaxation times were found not to affect the absorptance and THG conversion efficiency. We provide more details about this advantageous towards the practical implementation feature in the Supplemental Material [73].

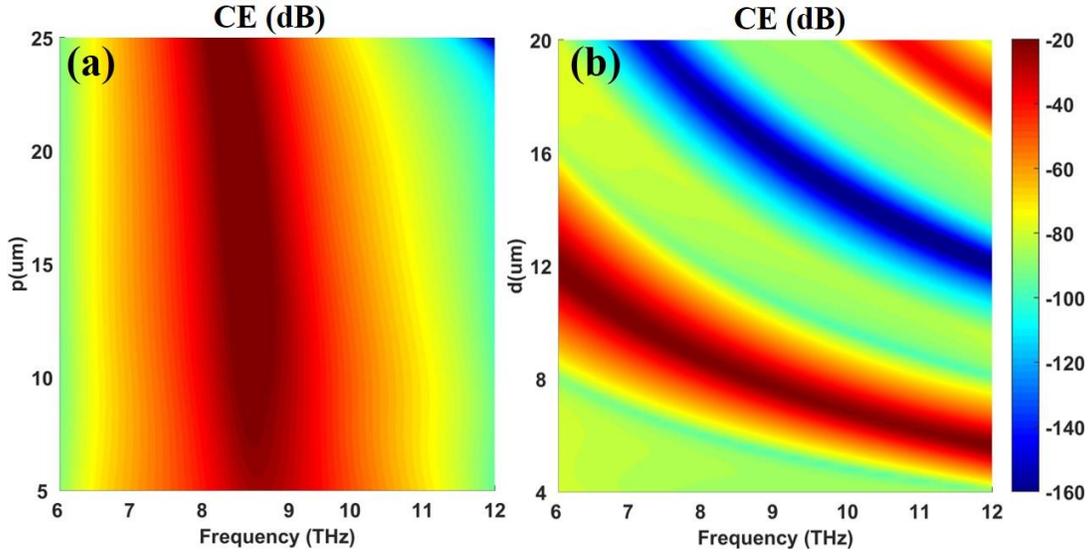

Figure 6 – THG CE contour plots of the proposed nonlinear hybrid structure at normal incidence as a function of the frequency and (a) the period or (b) height of the grating. The grating height is fixed to *d=8um* in (a) and the period is fixed to *p=8um* in (b).

Finally, the TH output power is computed with and without the graphene monolayer to prove that the addition of graphene is crucial in order to obtain enhanced nonlinear effects. The comparison results of the proposed hybrid structure and a similar structure made of a flat metallic substrate (no grating/ geometry shown in the inset of Fig. 7) with and without graphene on top are plotted in Fig. 7. In order to have a fair comparison, all the results are obtained under a varying incident angle wave and by using the same fundamental frequency of 8.8 THz. Moreover, the grating height and the substrate thickness are also kept identical during this comparison. The TH output power of the proposed graphene-covered hybrid grating [blue line in Fig. 7] has by far the highest value compared to all the other scenarios: i) the plasmonic grating with the same dimensions but without graphene on top [black line in Fig. 7], ii) the graphene-covered flat metallic substrate without the grating corrugations [green line in Fig. 7], and iii) the flat metallic substrate without the graphene monolayer on top [red line in Fig. 7]. *Under normal incidence, the TH radiation generated by the proposed graphene-covered plasmonic grating has an impressive twenty eight orders of magnitude THG enhancement compared to the same plasmonic grating but without graphene.* In a similar way, the case of the flat metallic substrate has much higher TH radiation when graphene is placed on top of it but still much lower values compared to the plasmonic grating case. Thus, it can be concluded that graphene plays a crucial role in the strong enhancement of the THG process. The same structure without the key nonlinear element of graphene will not produce such significant THG radiation. These results directly demonstrate the great potential of graphene in THz nonlinear optics. In addition, the THG process is much stronger in the case of the plasmonic grating configuration compared to the flat metallic substrate without corrugations since the grating structure can enable strong localized electric fields at its plasmonic resonance that strongly couple to graphene, as was mentioned before. Note that the THG output power remains relative insensitive across a broad incident angle range, especially between $[-30°, 30°]$, in agreement to the linear absorptance spectrum.

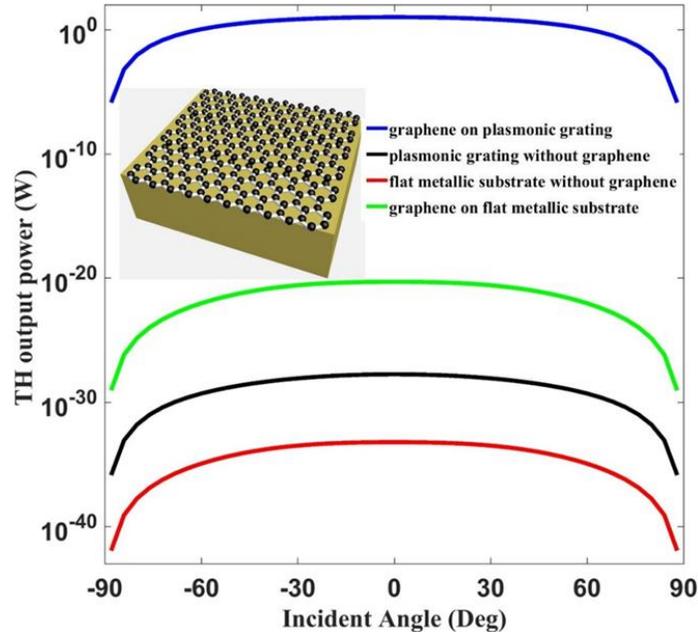

Figure 7 - Comparison of the TH output power of the proposed hybrid graphene-covered grating (blue line), the plasmonic grating without graphene (black line), the flat metallic substrate covered by graphene (green line/geometry shown in the inset), and the flat metallic substrate without graphene (red line). All the structures are excited by an 8.8 THz frequency incident wave under varying incident angles.

## IV. Four Wave Mixing

Another interesting third-order optical nonlinear procedure is FWM which typical requires very high input intensities to be excited. A feasible way to improve the efficiency of the FWM process is to increase the local field intensity at both input waves by using an artificially engineered structure [39, 83]. In the following, we demonstrate that the proposed graphene-covered grating can serve as an excellent platform to also boost this nonlinear process at THz frequencies. The strong coupling and interference between the plasmonic resonance of the metallic grating and the THz surface plasmon along the graphene monolayer can lead to strong local field enhancement, as was demonstrated before, which is expected to enhance FWM. During the FWM process, two $\omega_1$ and one $\omega_2$ photons are mixed and a photon is emitted at $\omega_3 = 2\omega_1 - \omega_2$. In order to take advantage of the strong field enhancement at the resonance and boost the FWM process, the two incident wave frequencies are chosen to be equal to $f_1 = 8.8 THz$ and $f_2 = 9.2 THz$. Thus, the generated FWM wave will be located at $f_3 = 8.4 THz$. The frequencies of the incident and generated waves are all very close to the maximum absorptance resonant frequency (8.8 THz). Hence, the electric fields induced by the incident and generated waves are expected to be greatly enhanced at these frequencies. The computed electric field distribution enhancement at the generated FWM wave $f_3 = 8.4 THz$ is demonstrated in the inset of Fig. 8(b).

Again, we use COMSOL to investigate the enhanced FWM nonlinear process based on the proposed nonlinear graphene-covered grating. The relevant schematic of this nonlinear procedure is illustrated in Fig. 8(a). The boundary conditions are the same with the THG simulations presented before except that one more electromagnetic wave solver is required to take into account the mixing mechanism introduced by the additional $\omega_2$ input wave. Both incident waves are TM polarized and have incident angles $\theta_1$ and $\theta_2$, respectively. In addition, both input intensities are selected to have low values equal to $20 kW/cm^2$. We measure the FWM radiated power through the upper boundary of the simulation domain by integrating the power outflow over the surface that surrounds the nonlinear structure, as it was performed before in the case of the THG process. The computed FWM output results are shown in Fig. 8(b). The incident angle $\theta_2$ is kept constant and equal to zero and the other incident angle $\theta_1$ varies from $0°$ to $\pm 90°$. The maximum output power is found to be 381 W at $\theta_1 = 0°$ and remains close to this peak value within a relative broad range of incident angles $(-30°, 30°)$. The FWM output power is symmetric with respect to $\theta_1$ and relatively insensitive to this incident angle. The same result is also obtained when the incident angles $\theta_1$ and $\theta_2$ of both input waves are simultaneously swept. The computed contour plot of this study is shown in Fig. 9. In this case, the used incident frequencies are the same with the results presented before in Fig. 8(b). It can be concluded that the FWM efficiency of the proposed graphene-covered grating is very high and relative insensitive to the excitation angles of both incident waves.

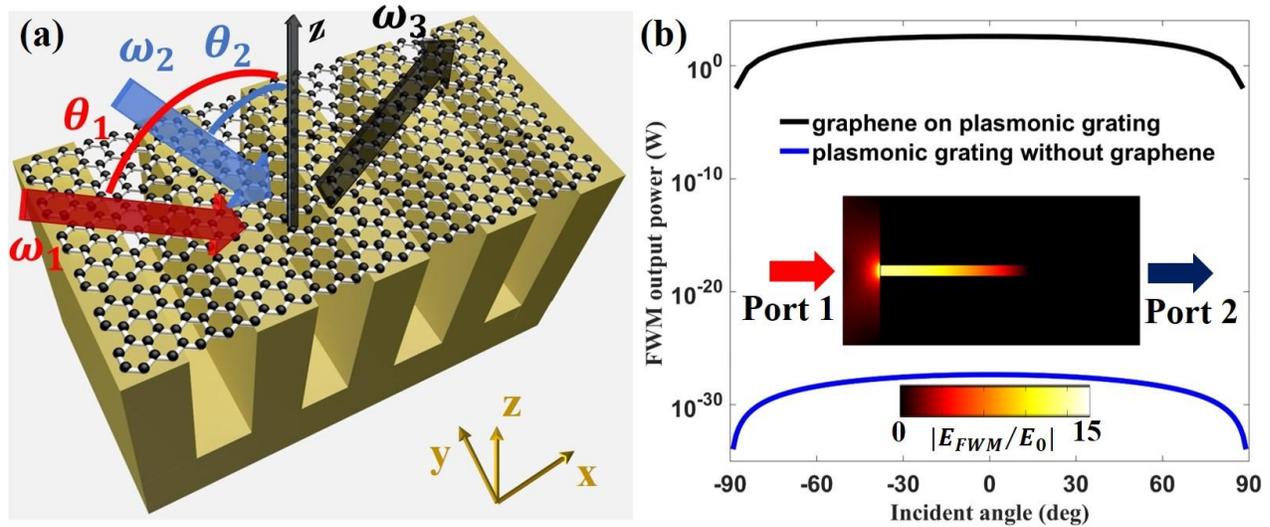

Figure 8 - (a) Schematic of the FWM process due to the nonlinear graphene-covered plasmonic grating. (b) Computed output power of the generated FWM wave $\omega_3$ as a function the excitation angle $\theta_1$ of the incident wave for the cases of the graphene-covered grating (black line) and only the grating without the graphene monolayer (blue line). Inset: The computed electric field enhancement distribution at the FWM frequency $f_3 = 8.4 THz$.

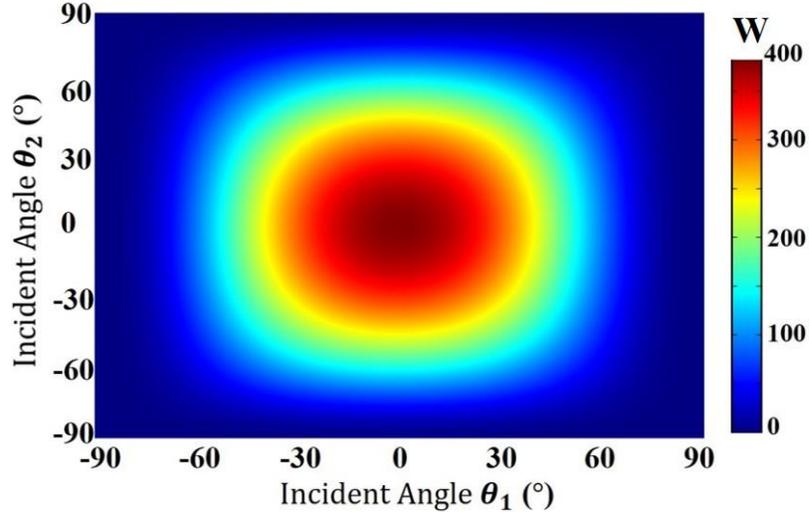

Figure 9 - Computed output power of the generated FMW wave as a function of the excitation angles of both incident waves impinging on the nonlinear hybrid graphene-covered grating.

The FWM efficiency can be calculated by computing the ratio $P_{out,tot}/P_{in,tot}$, where $P_{out,tot}$ and $P_{in,tot}$ are the total output and input powers, respectively. It is noteworthy that the FWM efficiency of the proposed hybrid grating can reach very high values of approximately 12% with a relatively low input intensity $20\,kW/cm^2$. To the best of our knowledge, such high nonlinear efficiency has never been reported before in the literature of both theoretical and experimental nonlinear plasmonic devices [38]. This efficiency is even higher compared to the THG process presented before because one more incident wave is contributing its power in the FWM process. The obtained high nonlinear efficiency is one of the major advantages of the proposed hybrid THz nonlinear graphene-plasmonic configuration. The FWM output power generated by the metallic grating without graphene on top is also calculated and plotted in Fig. 8(b) (blue line). *Clearly, the FWM output power is dramatically increased with the proposed graphene-covered grating by approximately twenty five orders of magnitude compared to a plain grating without graphene.* This comparison provides an additional proof of the key contribution of graphene in the boosted nonlinear response of the proposed plasmonic system.

As indicated before in Section III, the Fermi level $E_F$ will have a pronounced effect on the nonlinear response of the proposed structure. It will lead to different values in the nonlinear conductivity of graphene, which is given by Eq. (1). This effect is also predominant in the FWM process. The variation in the FWM output power due to increased $E_F$ is computed and shown in Fig. 10(a). The incident frequencies are again fixed to $f_1 = 8.8\,THz$ and $f_2 = 9.2\,THz$ in this case. The FWM output power is decreased by five orders of magnitude when the Fermi level is increased from 0.1 eV to 0.45 eV. This trend is consistent to the formula of the third-order nonlinear surface conductivity of graphene given by Eq. (1). Moreover, we explore the effect of the proposed hybrid structure's geometry on the FWM process, similar to our previous THG analysis. The results are shown in Figs. 10(b) and (c). The FWM output power is relative high when the period varies from 5 um to 25 um and reaches a maximum value of 1150 W for approximately 15 um period. This trend is similar to the THG CE contour plot shown before in Fig. 6(a), where the FF wave was fixed to 8.8 THz. The FWM output power changes dramatically with the grating's height, as demonstrated in Fig. 10(c). It reaches a maximum value of 382 W for *d=8um* and a minimum value of $2\times 10^{-11}W$ for *d=16um*. This trend is again similar to the THG CE contour plot shown in Fig. 6(b). The absorptance resonant frequency is strongly shifted when we change the grating height *d* [see Fig. 2(b)] and this leads to a dramatic variation in the FWM output power. Thus, it can be concluded that the output radiated power of the generated FWM wave can be tuned by either changing the graphene's Fermi level or the geometry of the proposed hybrid structure.

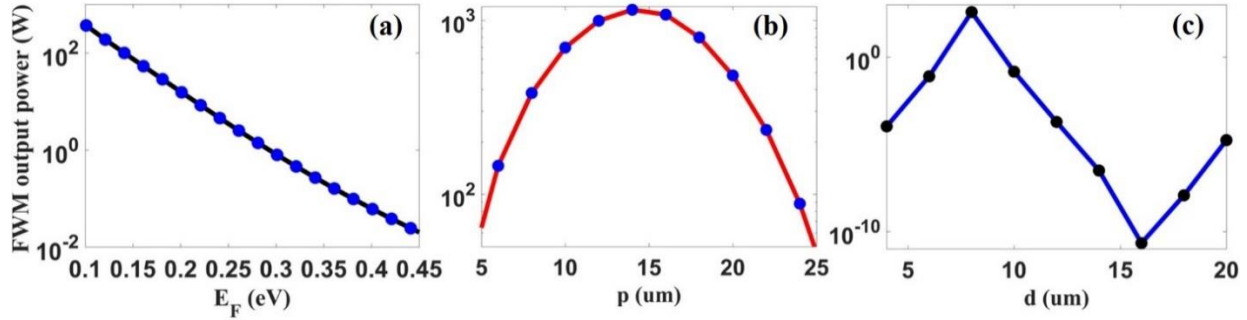

Figure 10 - Computed output power of the generated FWM wave by the nonlinear hybrid graphene-covered grating [schematically depicted in Fig. 8(a)] as a function of (a) the graphene's Fermi level with fixed structure dimensions equal to *p=8um, b=0.6um, d=8um*, (b) the grating period *p* when $E_F = 0.1 eV$ and the other dimensions are fixed to *b=0.6um* and *d=8um*, and (c) the grating height *d* when again $E_F = 0.1 eV$ and the other dimensions are fixed to *b=0.6um* and *p=8um*.

Finally, an alternative robust way to control the output radiation power of the generated FWM wave is achieved by varying the incident power of both excitation waves. The FMW signal is expected to follow a square power law behavior as a function of the input power $P_1$ of the first incident wave $\omega_1$ and a linear power law relation as a function of the input power $P_2$ of the second incident wave $\omega_2$ [33]. The effect of the input pump intensities $P_1$ and $P_2$ of the two incident waves with frequencies of $f_1 = 8.8 THz$ and $f_2 = 9.2 THz$, respectively, on the output power of the generated FWM wave is illustrated in Fig. 11. Indeed, the generated FWM power is approximately a square function of $P_1$ and a linear function of $P_2$. This result also demonstrates that $P_1$ has a stronger effect on the generated FWM power [39]. Hence, the generated FWM power can be the further enhanced by increasing the input power of both incident waves. Thermal damage to gold grating can occur only for very high input intensities in the order of $GW/cm^2$. In addition, graphene has an even higher melting point compared to gold and is not expected to be affected by thermal effects.

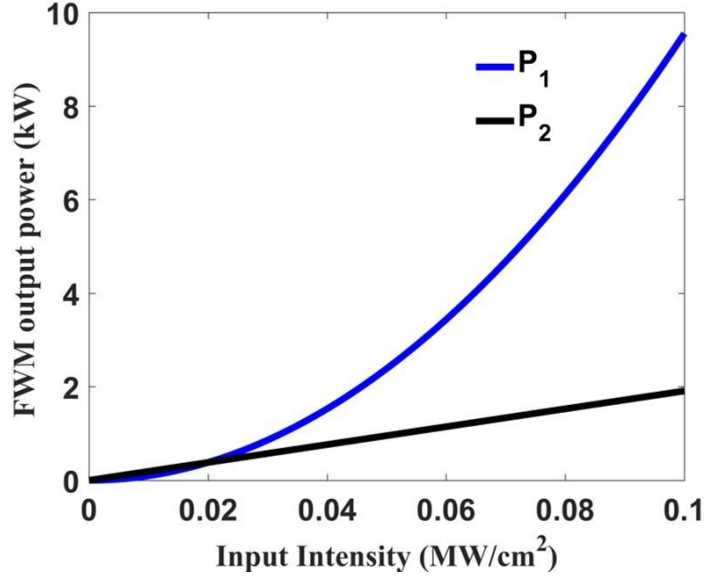

Figure 11 - The effect of the input pump intensities $P_1$ and $P_2$ on the output power of the generated FWM wave. The pump frequencies of the two incident waves are $f_1 = 8.8 THz$ and $f_2 = 9.2 THz$.

## V. Conclusions

In this work, we have analyzed and demonstrated enhanced nonlinear THz effects based on a new hybrid THz planar nonlinear device composed of a graphene monolayer placed over a metallic grating. The presented strong nonlinear response is mainly due to the localization and enhancement of the electric field at the absorption resonance of the proposed structure and the large nonlinear conductivity of graphene at low THz frequencies. It is demonstrated that the efficiency of both THG and FWM nonlinear processes can be dramatically enhanced by many orders of magnitude compared to other conventional non-hybrid metallic gratings and substrates. The presented nonlinear efficiencies are computed to be very large on the order of a few percent. They are higher compared to the majority of the state-of-the-art nonlinear planar plasmonic devices. Another major advantage of the proposed hybrid THz nonlinear configuration is that its nonlinear response can be dynamically tuned by using different mechanisms. In particular, it is presented that the output powers of both THG and FWM processes can be tuned by varying the metallic grating dimensions. This is due to the fact that the geometrical variations can lead to significant shifts in the absorptance resonant frequency, where the electric field, that triggers the nonlinear response, is greatly enhanced. Even more importantly, the nonlinear response can be dynamically modulated without altering the geometry of the proposed device and just by varying the graphene doping level. Finally, we demonstrate that the efficiencies of both THG and FWM processes can be further improved by increasing the input intensity of the incident waves.

The proposed graphene-covered metallic gratings can be realized with commonly used fabrication techniques. They can be built by a combination of chemical vapor deposition to efficient exfoliate graphene and the deposition of the graphene monolayer over a microscale gold grating which can be constructed by using conventional lithography techniques. The presented optimized grating design usually requires a relative high aspect ratio (groove depth $d$ over groove width $b$) with values between 10-20 to achieve strong absorption and nonlinear response but similar metallic gratings have recently been built based on photolithography [84], deep reactive-ion etching Bosch process [85], or nanoimprint lithography [86]. The proposed hybrid nonlinear graphene-plasmonic devices are envisioned to have several applications relevant to the new field of nonlinear optics based on 2D materials. They can be used in the design of THz frequency generators, all-optical signal processors, and wave mixers. Moreover, they are expected to be valuable components in the design of new nonlinear THz spectroscopy and noninvasive THz subwavelength imaging devices. Finally, the strong field confinement inside the nanoscale trenches and along graphene, achieved by the proposed hybrid grating, can be used to enhance dipole forbidden transitions on the atomic scale [87, 88].

## Acknowledgments
This work has been partially supported by the National Science Foundation (NSF) through the Nebraska Materials Research Science and Engineering Center (MRSEC) (grant No. DMR-1420645).

# Supplemental Material

# Hybrid Graphene-Plasmonic Gratings to Achieve Enhanced Nonlinear Effects at Terahertz Frequencies

Tianjing Guo, Boyuan Jin, and Christos Argyropoulos[*]

Department of Electrical and Computer Engineering, University of Nebraska-Lincoln, Lincoln, Nebraska 68588, USA

*christos.argyropoulos@unl.edu

## I. The effect of the graphene electron relaxation time

In the low terahertz (THz) frequency range, the linear conductivity of graphene can be expressed by the Drude formula: $\sigma_g = \frac{e^2 E_F}{\pi \hbar^2} \frac{j}{j\tau^{-1} - \omega}$ [S1], where $e$ is the electron charge, $E_F$ is the Fermi level or doping level of graphene, $\hbar$ is the reduced Planck's constant, $\omega = 2\pi f$ is the angular frequency, and $\tau$ is the electron relaxation time. Generally, the electron relaxation time characterizes the graphene loss and is described by the relation $\tau = \mu E_F / e v_F^2$ [S2, S3], where $\mu$ is the DC mobility of the graphene monolayer and $v_F = 1 \times 10^6 \, m/s$ is the Fermi velocity. The electron relaxation time can be affected by many factors, such as temperature, Fermi level (doping), external field, graphene sample quality, and the substrate material used [S4, S5]. Its value is in the picosecond range [S6, S7] because the mobility of graphene cannot experimentally exceed the upper limit $\mu = 20 \, m^2/(V \cdot s)$ at room temperature [S8, S9]. In the main paper, we assumed $\tau = 0.1 \, ps$ and graphene Fermi level $E_F = 0.1 \, eV$, which corresponds to a fixed low mobility of $\mu = 1 \, m^2/(V \cdot s)$.

In order to investigate the effect of graphene's electron relaxation time in the absorptance and nonlinear response of the proposed structure, we varied the relaxation time from $0.01 \, ps$ to $2 \, ps$, which corresponds to a mobility value ranging from $0.1 \, m^2/(V \cdot s)$ to $20 \, m^2/(V \cdot s)$, and computed the linear absorptance and third harmonic generation (THG) conversion efficiency (CE). The results are shown in Fig. S1. The input intensity used to compute the presented THG CE is low and equal to $20 \, kW/cm^2$. It can be seen in Fig. S1 that the resonant frequency of the proposed structure is always constant except for a minor redshift in both the absorptance and THG conversion efficiency for low electron relaxation time values. Furthermore, there is almost no change in their peak values with respect to the relaxation time. These results prove that there are no particular special requirements in the fabrication of the graphene monolayer used in the proposed hybrid grating structures, which is a major advantage towards their practical implementation.

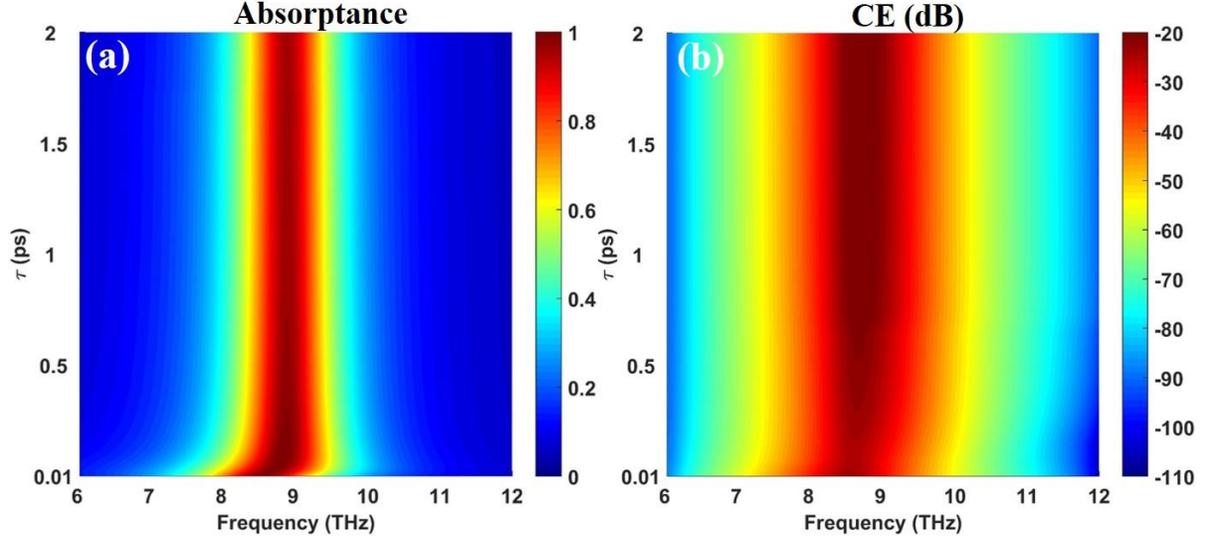

Figure S1 – (a) Computed absorptance and (b) THG CE contour plots in logarithmic scale at normal incidence as a function of the fundamental frequency and graphene's electron relaxation time.

## II.  Numerical method details

We employed the full-wave simulation software COMSOL Multiphysics to solve the linear and nonlinear Maxwell's equations and investigate the perfect absorption and enhancement of the nonlinear effects by the proposed graphene-covered plasmonic gratings.

**Geometry:** The proposed structure is composed of a gold grating covered by a graphene sheet, as shown in Fig. S2(a). This structure can be modeled as a two-dimensional (2D) periodic system since it is assumed to be infinite (or very large) along the y-direction. Periodic boundary conditions (PBC) are employed in both sides at the x-direction and port boundaries are placed up and down in the z-direction to create the incident plane wave. The PBC boundaries are indicated with green lines and the used ports 1 and 2 are shown by light blue lines in Fig. S2(b). The proposed structure is illuminated by a transverse magnetic (TM) polarized wave (the magnetic field is in the y-direction) with an incident angle $\theta$ with respect to the z-direction.

**Materials:** Graphene is modeled as a surface current due to its planar ultrathin 2D nature, which is represented by a red line in Fig. S2(b). The graphene's surface current is described by $J = \sigma_g E_{FF}$ and $J = \sigma_g E_{TH} + \sigma^{(3)} E_{FF}^3$ when it is used in the linear and nonlinear simulations, respectively. The $E_{FF}$ and $E_{TH}$ are the electric fields induced at the fundamental frequency (FF) and the third harmonic (TH) frequency, respectively. Note that these fields are monitored along the entire surface of graphene. The main electric field component that couples with graphene is the x-component [Fig. S4(c)] which is along the plane of the ultrathin graphene layer. The y-component [Fig. S4(d)] of the electric field does not couple to the zero-thickness graphene monolayer. The third-order nonlinear surface conductivity of graphene $\sigma^{(3)}$ is calculated by Eq. (1) in the main paper. The grating is made of gold with THz optical constants calculated by using the Drude model: $\varepsilon_{L,Au} = \varepsilon_\infty - f_p^2/f(f - i\gamma)$, where the parameters $f_p = 2069\,THz$, $\gamma = 17.65\,THz$ and $\varepsilon_\infty = 1.53$ are derived from fitting the experimental values [S10]. In the nonlinear simulations, we included the gold's nonlinear susceptibility $\chi_{Au}^{(3)} = 2.45 \times 10^{-19}\,m^2/V^2$. Hence, the Kerr third-order nonlinear permittivity of gold is given by: $\varepsilon_{NL,Au} = \varepsilon_{L,Au} + \chi_{Au}^{(3)} E_{FF}^2$, where $E_{FF}$ in this case is the induced field at the fundamental frequency monitored along the entire bulk gold volume (composed by both x- and y-components).

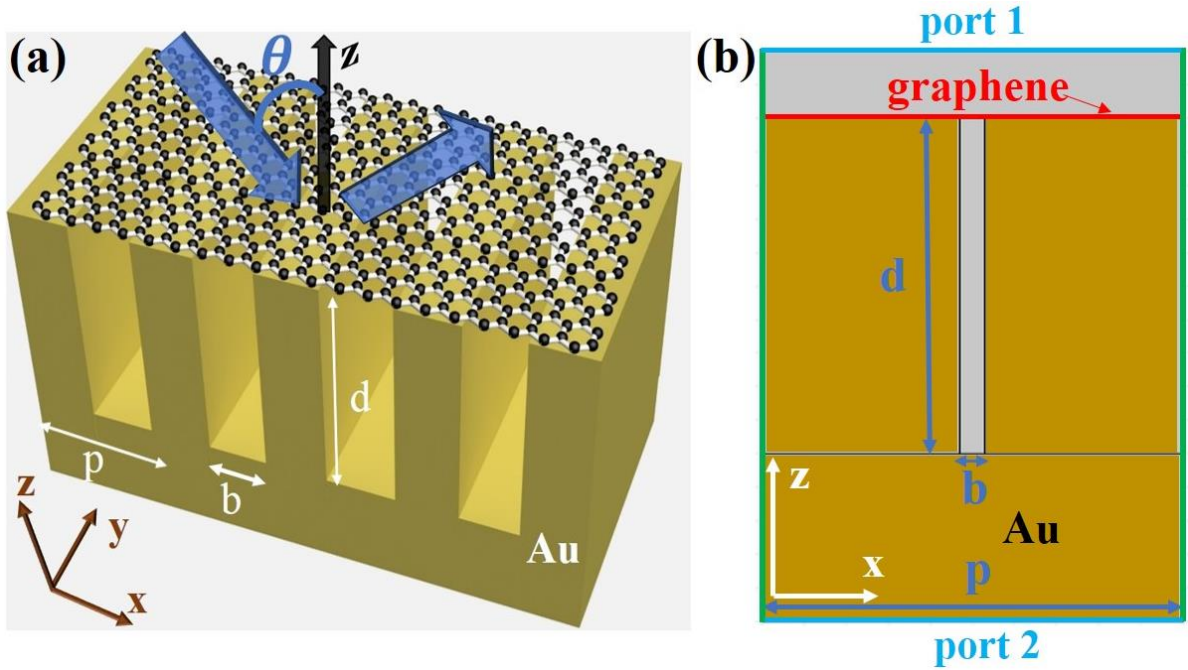

Figure S2 – (a) Schematic of the hybrid graphene-covered gold grating and (b) its simulation domain in COMSOL. The used PBC and port boundaries are shown with green and light blue lines in (b), respectively. The red line represents the graphene, which is modeled as surface current in the simulations. The yellow and gray areas represent gold and air, respectively.

**Results:** The S-parameter calculations were used in COMSOL to compute the total reflected and transmitted power flow through the input (port 1) and output (port 2) ports due to the proposed periodic structure. Their definitions in terms of the power flow are $S_{11} = \sqrt{\dfrac{power\ reflected\ back\ to\ port\,1}{emitted\ power\ from\ port\,1}}$ and $S_{21} = \sqrt{\dfrac{power\ transmitted\ to\ port\,2}{emitted\ power\ from\ port\,1}}$. The power delivered to each port (either by reflection or transmission) is calculated by

$P = \int_C \vec{S} \cdot \vec{n} = \frac{1}{2} \int_C \text{Re}\{\vec{E} \times \vec{H}^*\} \cdot \vec{n}$ , where $\vec{S}$ is the time-averaged Poynting vector, C is the boundary curve of each port, $\vec{n}$ is the boundary norm vector, $\vec{E}$ is the electric field vector, and $\vec{H}$ is the magnetic field vector along the ports. The emitted power from port 1 has a fixed value of 1600W, which corresponds to the low intensity value of $20\,kW/cm^2$. This value is defined during our COMSOL simulations and is varied for the nonlinear simulations. The reflection coefficient of the hybrid periodic grating is equal to $\Gamma = S_{11}$ and the transmission coefficient is $T = S_{21}$. The squared amplitudes of the reflection and transmission coefficient values give the reflectance $|\Gamma|^2$ and transmittance $|T|^2$ of the proposed system. The total absorptance can be calculated by the formula $A = 1 - |\Gamma|^2 - |T|^2$. In the current grating configuration the transmittance is equal to zero due to the thick gold substrate and the absorptance is computed by the simpler relationship $A = 1 - |\Gamma|^2$ [S3, S11-S14], since in the current grating configuration the transmittance is equal to zero.

Only one electromagnetic wave solver is used to perform the linear absorptance simulations at the fundamental frequency (FF). In order to compute the THG nonlinear radiation, an additional electromagnetic wave solver needs to be included and coupled to the FF solver. The harmonic wave coupling frequency domain method is used in our calculations by nonlinearly connecting several physics interfaces that model the structure in each frequency of the proposed nonlinear process. In the case of THG simulations, COMSOL will solve the nonlinear Maxwell's equations at the TH frequency $f_{TH} = 3f$, where $f$ is the fundamental frequency. In this case, a wave with frequency $3f$ will be generated, when an incident wave with FF $f$ excites the proposed nonlinear graphene-covered grating. The radiated output power of the TH wave is computed by calculating the integral $\int_C \vec{S} \cdot \vec{n}$ over the surface of the structure, where $\vec{S}$ is the Poynting vector crossing the boundary surface C and $\vec{n}$ is the boundary norm vector. Here, the boundary surface C is the top port boundary which is shown by the light blue line in Fig. S2(b). The grating does not reflect or leak any additional high-order diffraction modes or surface waves and the generated reflected nonlinear radiation is passing only from the top port boundary, which is placed in the far field.

During the four wave mixing (FWM) process, two waves with frequency $f_1$ and one wave with frequency $f_2$ are mixed and a new wave is generated and emitted with frequency $f_3 = 2f_1 - f_2$. Thus, three electromagnetic wave solvers are needed to take into account the FWM mechanism and compute the emitted $f_3$ wave. The first two solvers with frequencies $f_1$ and $f_2$ would operate as input waves to the nonlinear third solver that computes the emitted wave with frequency $f_3$. Both input waves are TM polarized and have incident angles $\theta_1$ and $\theta_2$, respectively. In addition, they have the same input intensities. The same method with the presented in the previous paragraph

THG radiation calculation is used to compute the generated FWM output power. Finally, the enhanced electric field distribution of each process is calculated by the ratio $|E/E_0|$, where $E$ is the electric field at any point in the computational domain that can be calculated in different frequencies.

### III. Response of flat gold substrate with and without graphene

Surface plasmons excited at bounded dielectric-metal geometries are called localized surface plasmons (LSP) [S15]. They have been reported to play an important role to enhance absorptance in metallic gratings [S16-S18]. We have also verified this property by analytically calculating the absorptance of an optimized gold grating under normal incidence, which is shown by the green line in Fig. 1(b) in the main paper. Moreover, high enhancement of the electric field is obtained at the LSP resonance frequency (see inset of Fig. 1(b) in the main paper) that can be utilized to boost the total absorptance of graphene [S17] or to enhance different nonlinear effects [S19]. Note that surface plasmon polaritons (SPP), instead of LSP, can be excited along a flat metallic film [S15]. In the case of the flat gold substrate, the dielectric adjacent to the gold surface is air and, as a result, SPP cannot be excited in the currently studied low THz frequency range [S15]. Hence, the absorptance remains very low along these frequencies, which is calculated and demonstrated by the blue line in Fig. S3. The flat gold substrate acts as an almost perfect mirror to the THz radiation and most of the incident light is reflected back leading to very small absorptance due to the large reflection coefficient and, as a result, reflectance. This is in sharp contrast to the obtained absorptance enhancement in our proposed hybrid graphene-covered gold grating which comes from the strong coupling between the LSP along the gold grating and the SPP in the graphene sheet. The brown line in Fig. S3 represents the absorptance of graphene placed over the flat gold substrate with a schematic of this geometry shown in the inset of Fig. S3. The absorptance of this configuration is also low and almost identical to the case without graphene (blue line in Fig. S3) because again LSP or SPP cannot be excited. The low absorptance of the pure flat gold substrate and planar gold with graphene cases naturally leads to low nonlinear efficiency. This is verified by Fig. 7 in the main paper, where the THG conversion efficiency of the planar gold substrate with (green line) and without (red line) graphene on top is plotted.

### IV. Power flow and perfect absorption considerations

To further demonstrate the perfect absorption obtained by the proposed hybrid graphene-plasmonic gratings, we calculate the power flow versus frequency at the interface between air and the grating's gold material by integrating the time-averaged Poynting vector given by:

$P = \int_C \vec{S} \cdot \vec{n} = \frac{1}{2} \int_C \text{Re}\{\vec{E} \times \vec{H}^*\} \cdot \vec{n}$, where $\vec{S}$ is the time-averaged Poynting vector, C is the boundary curve at the gold/air interface extended along the gold part of the grating and not along the air upper part of the corrugations, $\vec{n}$ is the boundary norm vector along this interface, $\vec{E}$ is the electric field vector and $\vec{H}$ is the magnetic field vector. The skin depth of gold around the presented perfect absorption resonance frequency (8.8 THz) is approximately $\delta = 0.02um$ [S20, S21] and no power is expected to penetrate inside gold below this ultrathin skin depth. We calculate the power going into gold versus frequency by integrating the time-averaged Poynting vector over three boundary curves along different depths inside the gold part of the grating that are shown by the red, blue, and green lines in the inset of Fig. S4(a). Note that these lines are only extended along the gold part of the grating and not along the upper air part of the corrugation trenches. The red line represents exactly the interface between air and gold, the green line is placed at approximately the skin depth of gold (0.02um), and the blue line is located deep inside gold and five times larger compared to the gold's skin depth (0.1um). We integrated the time-averaged Poynting vector over the curves of interest to calculate the total power going inside gold in these three different cases versus the incident radiation frequency. The computed power for these three cases is shown in Fig. S4(a), where it is demonstrated that the power flow is much smaller and almost zero deep inside gold (five times larger compared to the gold's skin depth) (blue line) compared to the power at the interface (red line).

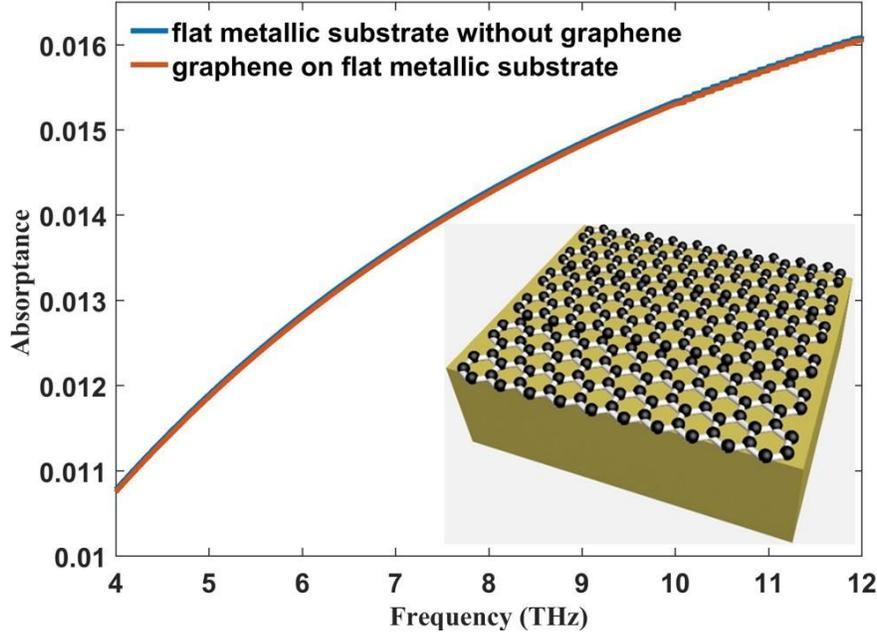

Figure S3 – Comparison of the computed absorptance between the pure gold substrate and planar gold substrate with graphene cases. Inset: A schematic of the flat gold substrate covered by graphene.

The minimum in the power flow spectrum obtained at the absorption resonance frequency (8.8THz) in Fig. S4(a) for values less than the gold's skin depth proves that most power goes into the grating's trenches and not inside gold at this frequency point, as it is also shown by the time-averaged power arrows plotted at the resonance in the close-up of Fig. S4(b). The power can fully penetrate and interact with the ultrathin 2D graphene placed on top of the gold grating corrugations, and this can be further derived by observing Fig. S4(b), where the computed absolute values of the electric field enhancement distribution and the arrows representing the power flow are plotted at the resonance frequency. It can be clearly seen in Fig. S4(b) that all the incident power is fully absorbed inside the elongated trenches of the metallic grating, since the power flow arrows become smaller as they travel deeper inside them. The majority of the field and power enhancement happens exactly at the interface between the grating trenches and air, where graphene is located. Hence, the power of the incident radiation, which is fully absorbed into the corrugations' trenches, is strongly interacting with graphene that leads to the enhanced nonlinear response presented in the paper. In addition, the x- and y-components of the real part of the electric field enhancement distribution at the resonance frequency are shown in Figs. S4(c) and (d), respectively, to further prove the aforementioned point. The absolute value of the electric field shown in Fig. S4(b) is computed by the formula $|E| = \sqrt{|E_x|^2 + |E_y|^2}$.

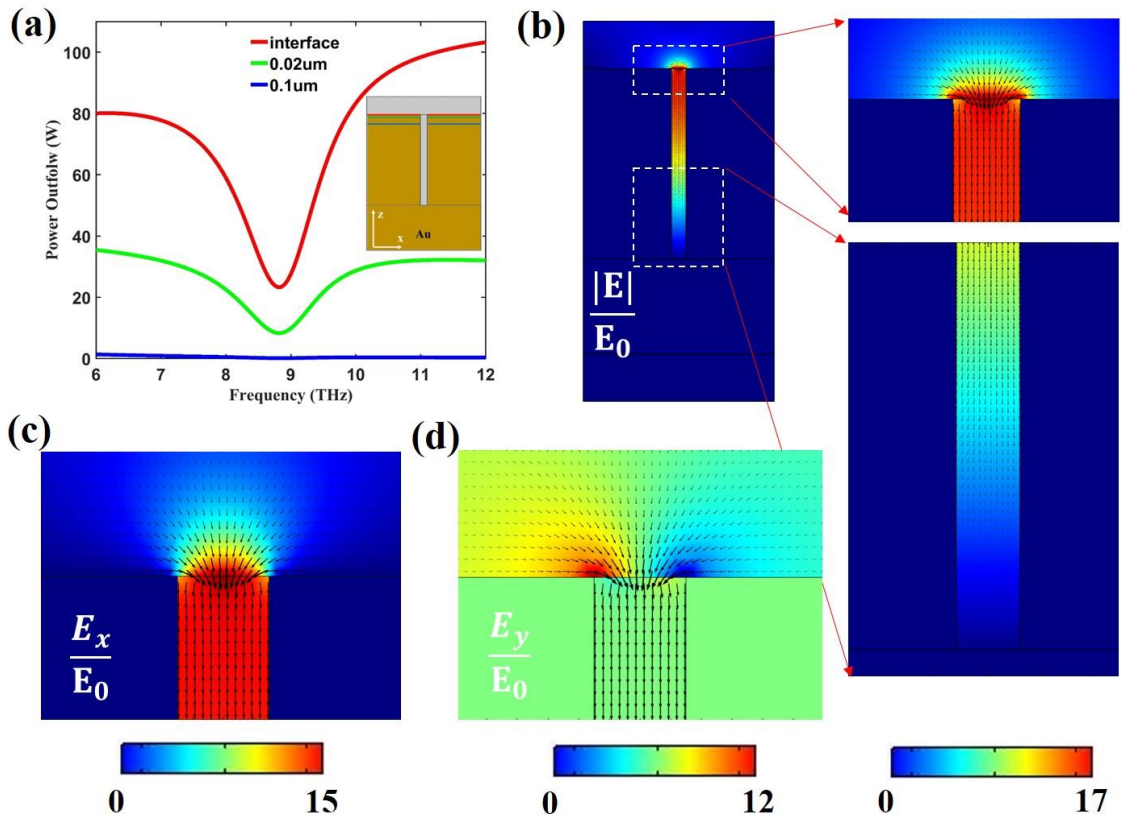

Figure S4 – (a) Computed power flow at the interface between air and gold (red line), almost equal to the gold's skin depth (green line), and four times larger (blue line) than the gold's skin depth. The inset shows a schematic with the corresponding positions where the power is calculated. (b) The computed absolute value of the electric field enhancement distribution and the power flow arrows inside the structure at the resonance frequency. The power flow

is dissipated and fully absorbed inside the air corrugations' trenches. (c), (d) Real part of x- and y-components of electric field enhancement distribution and the power flow arrows at the resonance frequency. The results are obtained for grating parameters p=8um, b=0.6um, d=8um. The graphene's Fermi level $E_F$ is equal to $0.1eV$.

It is obvious from Fig. S4 that all the energy is absorbed inside the corrugations and there are no surface waves travelling along the grating. To further prove this point, we create two new simulation models to clearly demonstrate that all the power is absorbed inside the trenches and there are no surface waves. In the first model, we consider a finite number of corrugations to simulate a more realistic situation where 16 corrugations are used and a continuous graphene sheet is placed on top of them. Perfect electric conductor (PEC) boundary conditions are employed in both sides at the x-direction and port boundaries are placed up and down in the z-direction to create the incident plane wave. The computed absorptance shown in Fig. S5(a) perfectly matches the single unit cell surrounded by PBCs simulation results presented in the main paper. Note that there is a small difference in the absorption resonance frequency (9.1THz) in this case compared to the infinite structure due to the minor detuning introduced by the finite structure of Fig. S5(a) and the slightly different grating parameters used. The black arrows in Fig. S5(b) depict the power flow of the finite structure at the absorption resonance frequency and clearly demonstrate that all the power is absorbed inside the corrugation trenches, similar to the results presented in Fig. S4. The power is not reflected back and it is fully absorbed as it travels along the trenches. This further proves the predicted perfect absorption response at the resonance frequency of the proposed hybrid grating and provides a clear evidence that there are no additional surface waves travelling along its interface.

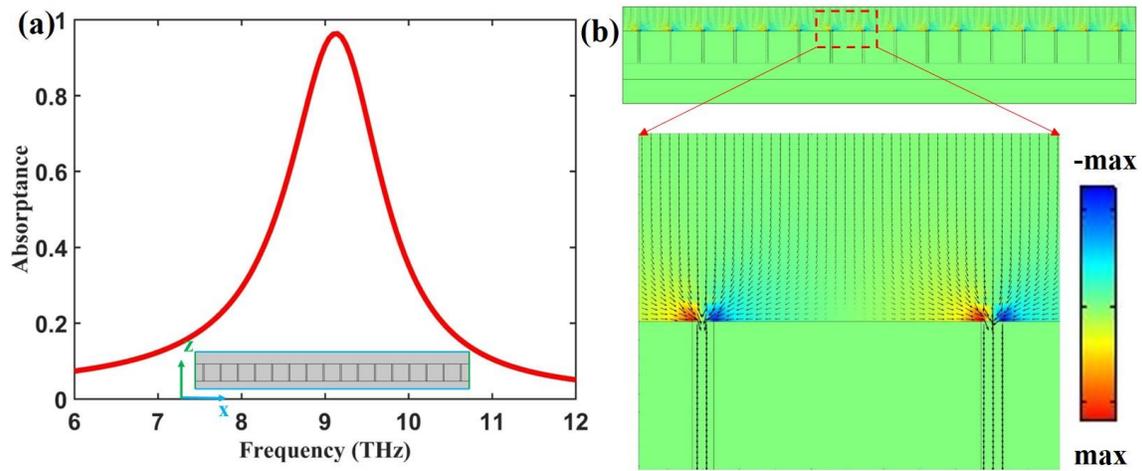

Figure S5 – (a) Computed absorptance spectra of the proposed hybrid grating made by a finite number of corrugations. The inset represents the simulated structure. (b) The y-component of the real part of the electric field enhancement distribution normalized at its maximum value at the absorption resonance frequency and the flowing of the power (depicted by arrows) inside the structure. The results are obtained for grating parameters p=8um, b=0.6um, d=8um. The graphene's Fermi level $E_F$ is equal to $0.1eV$.

In the second model, we employ a different way to calculate the total absorbed energy by integrating the total power dissipation density over the volume of the structure. This type of absorption energy calculations are usually performed in scattering problems and not in reflection/transmission problems similar to the currently proposed hybrid grating configuration. The total power dissipation density is calculated by using the formula $\vec{E}\cdot\vec{J}^*$, which can be obtained by the COMSOL predefined variable $Qh$. Thus, the total absorbed power of the finite hybrid grating is computed by $P_{abs} = \frac{1}{2}\int_S \mathrm{Re}\{\vec{J}^*\cdot\vec{E}\}$ [S22], where $J$ is the current density, S represents the surface of the proposed structure since our model is two-dimensional. By integrating the total power dissipation density over the blue region in Fig. S6(a), which includes both the lossy gold and graphene materials, we obtain the total dissipation power of the hybrid grating. In the finite grating simulation presented in Fig. S6(a), the background electric field formalism is employed to create the normal incident plane wave and perfect matched layers (PML) are used to fully absorb all outgoing waves scattered by the grating. The computed total dissipation energy by the proposed structure is shown in Fig. S6(b), where it is obvious that the trend and peak position of this curve are similar with the previous absorptance result shown in Fig. S5(a) for an identical structure but obtained by using the reflectance and transmittance values. The absorbed power reaches to a maximum value at the same resonance frequency and then drops to very low values, as it is expected and also predicted in the main paper. Thus, we can conclude that the total dissipation energy result computed in Fig. S6 matches almost perfectly the absorptance results computed by the models in Fig. S5 and the results **(Fig. 3)** presented in the main paper.

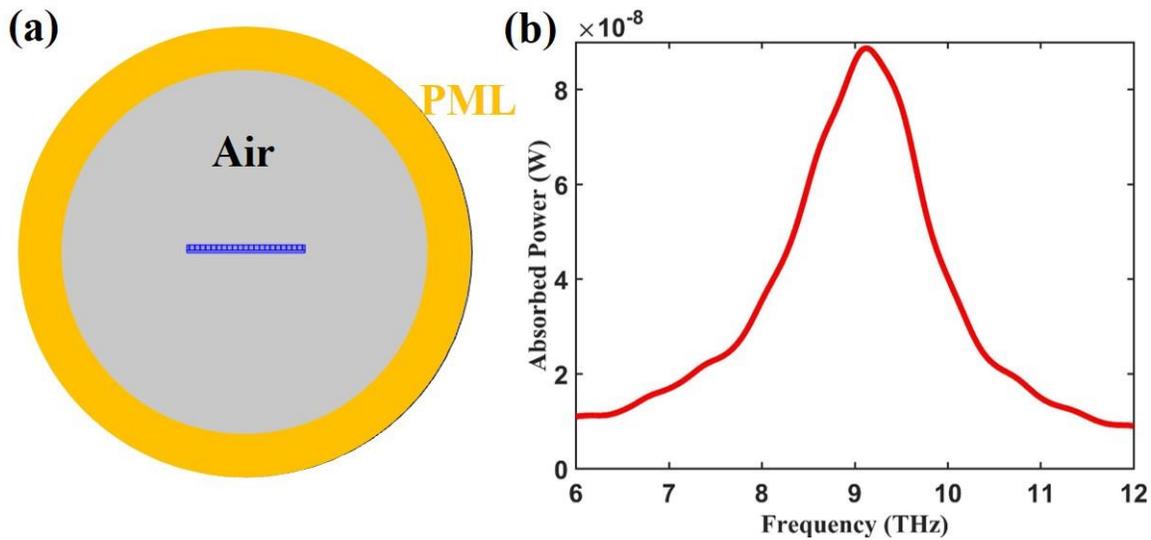

Figure S6 – (a) Scattering and absorption model of the finite hybrid graphene-covered grating. (b) The calculated over a broad frequency range total dissipation energy along the hybrid grating. The computed absorption peak is identical to the absorptance peak shown in Fig. S5(a) and the main paper. The results are obtained for grating parameters p=8um, b=0.6um, d=8um. The graphene's Fermi level $E_F$ is equal to $0.1eV$.